\documentclass[conference]{IEEEtran}

\usepackage{cite}       

\usepackage{amsmath}    
\usepackage{amssymb}    
\usepackage{amsfonts}   
\newcommand{\Si}{\Gamma_n}              
\newcommand{\Ef}{\Gamma^*_n}            
\newcommand{\Cef}{\overline{\Gamma}^*_n}     
\newcommand{\Asi}{\Phi_n}
\newcommand{\Aef}{\Phi^*_n}             
\newcommand{\Caef}{\overline{\Phi}^*_n}
\newcommand{\Sd}{\Lambda_n}             
\newcommand{\Sda}{\Lambda^*_n}
\newcommand{\Csda}{\overline{\Lambda}^*_n}
\newcommand{\Msd}{\mathbf{g}}
\newcommand{\SetR}{\mathbb{R}}

\newcommand{\condmid}{\,|\,} 
\newcommand{\GF}{\mathrm{GF}}
\newtheorem{definition}{Definition}
\newtheorem{theorem}{Theorem}

\begin{document}

\title{Average Entropy Functions}

\author{\authorblockN{Qi Chen, Chen He, Lingge Jiang, Qingchuan Wang}
\authorblockA{Dept. of Electronic Eng.\\
Shanghai Jiao Tong Univ.\\
Shanghai, China 200240\\
Email: \{cq094, chenhe, lgjiang, r6144\}@sjtu.edu.cn}}


%


\maketitle

\begin{abstract}
THIS PAPER IS ELIGIBLE FOR THE STUDENT PAPER AWARD. The closure of
the set of entropy functions associated with $n$ discrete variables,
$\Cef$, is a convex cone in $(2^n-1)$-dimensional space, but its
full characterization remains an open problem. In this paper, we map
$\Cef$ to an $n$-dimensional region $\Caef$ by averaging the joint
entropies with the same number of variables, and show that the
simpler $\Caef$ can be characterized solely by the Shannon-type
information inequalities.
\end{abstract}


%

\section{Introduction}
Given an $n$-dimensional discrete random vector
$\mathbf{X}=(X_{1},\ldots,X_{n})$, for each non-empty subset
$\alpha$ of $\mathcal{N}= \{1,2, \ldots,n\}$ there is a joint
entropy $H(X_{\alpha})$ with $X_{\alpha} = (X_{i})_{i\in\alpha}$, and
the $2^n-1$ joint entropies form the entropy function
$(H(X_{\alpha}))_{\alpha\subseteq\mathcal{N}, \alpha\ne\emptyset}$ of
$\mathbf{X}$.  We can then define $\Ef \subseteq \SetR^{2^n-1}$ as the
set of all possible entropy functions involving $n$
discrete random variables, and $\Cef$ as its closure.  A vector
$\mathbf{H} \in \SetR^{2^n-1}$ is called entropic if $\mathbf{H} \in
\Ef$, and almost entropic if $\mathbf{H} \in \Cef$ \cite{Y97}.

All $\mathbf{H}=(H_\alpha)_{\alpha\subset{\mathcal{N}},\alpha\neq\emptyset}\in\Cef$
satisfy the following Shannon-type information inequalities for any
subsets $\alpha$, $\beta$ of $\mathcal{N}$ (we let $H_{\emptyset} = 0$
for convenience):
\begin{gather}
\label{eq:shannon1}
H_\alpha\geq0, \\
\label{eq:shannon2}
H_\alpha\leq H_\beta,\quad \alpha\subseteq\beta, \\
\label{eq:shannon3}
H_\alpha+H_\beta \geq H_{(\alpha\cup\beta)}+H_{(\alpha\cap\beta)}.
\end{gather}
However, \eqref{eq:shannon1}--\eqref{eq:shannon3} are not sufficient
conditions for an $\mathbf{H} \in \SetR^{2^n-1}$ to be almost entropic
when $n \ge 4$ \cite{ZY98}.  In other words, denoting by $\Si$ the set of vectors in
$\SetR^{2^n-1}$ satisfying \eqref{eq:shannon1}--\eqref{eq:shannon3},
we have
\begin{equation}
\Cef\subsetneq\Si,\quad n\geq4.
\end{equation}
A number of non-Shannon-type information inequalities satisfied by the
members of $\Cef$ have subsequently been found in
\cite{ZY98,YZ01,DFZ06}, but the full characterization of $\Cef$ remains
an open problem.

In this paper, we will show that an averaged version of $\Cef$ can be
more easily characterized.

\begin{definition}
For a vector
$\mathbf{H}=(H_\alpha)_{\alpha\subset{\mathcal{N}},\alpha\neq\emptyset}\in\mathbb{R}^{2^{n}-1}$,
we define its average as
\begin{equation}
\Psi(\mathbf{H})\triangleq(h_1,\ldots,h_n),
\end{equation}
where $h_k={\binom{n}{k}}^{-1}\sum_{\mid\alpha\mid=k}H_\alpha$.
If $\mathbf{H}$ is the entropy function of random vector $\mathbf{X}$,
we call $\mathbf{h}=\Psi(\mathbf{H})$ the average entropy function.  $\Psi$ then
maps $\Ef$ to the set $\Aef\triangleq\Psi(\Ef)$ of all average entropy
functions, $\Cef$ to the closure $\Caef$, and $\Si$ to $\Asi\triangleq\Psi(\Si)$.
\end{definition}

From the definition \eqref{eq:shannon1}--\eqref{eq:shannon3} of $\Si$,
$\Asi$ can be given by
\begin{equation}
\label{eq:Asi}
\begin{split}
  \Asi=\{(h_1,\ldots,h_n)\condmid & h_{k-1}-2h_{k}+h_{k+1}\leq0,\\
  &k=1,\ldots,n\},
\end{split}
\end{equation}
where we let $h_{0}=0$ and $h_{n+1}=h_{n}$ for convenience.  $\Caef$
is obviously a subset of $\Asi$ since $\Cef\subseteq\Si$, but we will
show that they are actually equal.  In other words, $\Caef$ is
characterizable solely with the Shannon-type information inequalities.

\vspace{2mm}
\begin{theorem}
  $\Caef=\Asi$.
\end{theorem}
\vspace{2mm}

This theorem will be proved in the next section.

\section{Proof of the Theorem}
It is only necessary to prove that $\Asi\subseteq\Caef$.  We first
introduce a one-to-one transform to give $\Asi$ a simpler form.

\begin{definition}
For a vector
$\mathbf{h}=(h_1,\ldots,h_n)\in\mathbb{R}^n$, we define its second-order
difference as
\begin{equation}
\Theta(\mathbf{h})=(g_1,\ldots,g_n),
\end{equation}
where $g_k=h_{k-1}-2h_{k}+h_{k+1}$, $k=1,\ldots,n$, with
$h_{0}=0$ and $h_{n+1}=h_{n}$.  $\Theta$ maps $\Aef$ to
$\Sda\triangleq\Theta(\Aef)$, $\Caef$ to $\Csda$, and $\Asi$ to $\Sd
\triangleq \Theta(\Asi)$.
\end{definition}

From \eqref{eq:Asi}, we have
\begin{equation}
\Sd=\{(g_1,\ldots,g_n)\condmid g_k\leq 0,\ k=1,\ldots,n\}.
\end{equation}

As $\Psi$ and $\Theta$ are both linear maps, and $\Cef$ is a convex
cone \cite{ZY97}, $\Caef$ and $\Csda$ are both convex cones as well.
Therefore, to prove that $\Asi\subseteq\Caef$ or equivalently
$\Sd\subseteq\Csda$, it is sufficient to prove that
\begin{equation}
\Msd_k\triangleq(\underbrace{0,\ldots,0}_{k-1},-a,0,\ldots,0)\in\Sda
\end{equation}
for  $k=1,\ldots,n$ and some $a>0$. In other words, for each $k$ we
need to find a random vector $\mathbf{X}$ whose
average entropy function is
\begin{equation}\label{eq:hk}
\mathbf{h}_k\triangleq\Theta^{-1}(\Msd_k)=a\cdot(1,2,\ldots,k,\ldots,k).
\end{equation}
This $\mathbf{X}$ can be constructed from a Reed-Solomon code.
Specifically, let $q$ be a power-of-two larger than $n$,
$\mathcal{C}$ be the codeword set of an $(n,k)$ Reed-Solomon code on $\GF(q)$, and
$\mathbf{X} = (X_{1},\ldots,X_{n})$ be a random codeword uniformly
distributed over $\mathcal{C}$, then the entropy function of
$\mathbf{X}$ is \eqref{eq:hk} with $a = \log q$, as shown below.

Let $j_1,\ldots,j_n$ be distinct indices in ${1,\ldots,n}$.
According to the properties of Reed-Solomon codes, given any
$x^*_{j_1},\ldots,x^*_{j_k}\in \GF(q)$, there exists a unique
$\mathbf{x}=(x_1,\ldots,x_n)\in\mathcal{C}$ with $x_{j_l} =
x^*_{j_l}$, $l=1,\dotsc,k$.  For any $x^*_{j_1}\in\GF(q)$, there are
thus $q^{k-1}$ codewords $\mathbf{x}\in\mathcal{C}$ with $x_{j_1} =
x^*_{j_1}$, one for each value combination on $k-1$ other positions,
and since $\mathbf{X}$ is equal to each codeword with probability
$q^{-k}$, we have $p(X_{j_1}=x^*_{j_1})=q^{-1}$, so
$H(X_{j_1})=\log{q}$. Similarly, $H(X_{j_1},X_{j_2})=2\log{q}$,
\ldots, $H(X_{j_1},\ldots,X_{j_k})=k\log{q}$. For $l=k+1,\ldots,n$,
given $x_{j_1},\ldots,x_{j_l}$, there is either one matching
codeword in $\mathcal{C}$ or none, therefore
$p(X_{j_1}=x_{j_1},\ldots,X_{j_l}=x_{j_l})$ is $q^{-k}$ on its
support, and $H(X_{j_1},\ldots,X_{j_l})=k\log{q}$.  Consequently,
the average entropy function of $\mathbf{X}$ is \eqref{eq:hk} with
$a = \log q$ as desired, and for each $l$, all $\binom{n}{l}$
$l$-variable joint entropies of $\mathbf{X}$ that are being averaged
actually have the same value.\hfill\QED

\section{Discussion}
Determination of $\Cef$ is important due to its close connection to
the capacity region of general multi-source multi-sink wired networks
\cite{XYZ07,duality-entropy-func-network-codes}, but this seems to be
a difficult problem, and even if a full characterization is found,
computational difficulties due to $\Cef$'s high dimensionality and
complex structure might reduce its usefulness in practice
\cite{inf-many-info-ineq}.  What we have shown is that the region
$\Caef$ obtained by averaging the $k$-variable joint entropies has a
much simpler structure: it is not affected by the non-Shannon
information inequalities, and the linear Reed-Solomon codes used in
the proof suggest that the suboptimality of linear network coding is
also hidden by this averaging.  On one hand, this means that further
work on the characterization of $\Cef$ must focus on the variation
among the $k$-variable entropies, not just their averages.  On
the other hand, many practically interesting networks have a somewhat
symmetric structure, possibly in a statistical sense, and an
appropriately averaged version of $\Cef$ (not necessarily as
simplistic as $\Caef$) might provide a tractable method for the
determination of their capacity regions.

Average entropy functions are also closely related to the MAP EXIT
functions discussed in e.g.\ \cite{maxwell-constr} for large $n$.


\section*{Acknowledgment}
This paper was supported by National Natural Science Foundation of
China Grants No.~60772100 and No.~60872017.




\begin{thebibliography}{1}
\bibitem{Y97}
R.~W. Yeung, ``A framework for linear information inequalities,''
\emph{{IEEE} Trans. Inf. Theory}, vol.~43, pp.~1924--1934, Nov. 1997.
\bibitem{ZY98}
Z.~Zhang and R.~W. Yeung, ``On characterization of entropy function
via information inequalities,'' \emph{IEEE Trans. Inf. Theory}, vol.~44, pp.~1440--1452, Nov. 1998.
\bibitem{YZ01}
X.~Yan, R.~Yeung and Z.~Zhang, ``A class of non-Shannon type
information inequalities and their applications,'' \emph{IEEE Int. Symp. Inf. Theory},
  Washington, DC, June 2001.
\bibitem{DFZ06}
R.~Doughter, C.~Freiling and K.~Zeger, ``Six new non-Shannon
information inequalities,'' \emph{IEEE Int. Symp. Inf. Theory}, Seattle, WA, June
  2006.
\bibitem{ZY97}
Z.~Zhang and R.~W. Yeung, ``A non-Shannon type conditional inequality
of information quantities,'' \emph{IEEE Trans. Inf. Theory}, vol.~43, pp.~1982--1986, Nov. 1997.
\bibitem{XYZ07}
X.~Yan, R.~Yeung and Z.~Zhang, ``The capacity region for multi-source
multi-sink network coding,'' \emph{IEEE Int. Symp. Inf. Theory}, Nice,
France, June 2007.
\bibitem{duality-entropy-func-network-codes}
T.~Chan and A.~Grant, ``Dualities between entropy functions and
network codes,'' \emph{IEEE Trans. Inf. Theory}, vol.~54, no.~10,
pp.~4470--4487, Oct. 2008.
\bibitem{inf-many-info-ineq}
F.~Mat\'u\v{s}, ``Infinitely many information inequalities,'' \emph{IEEE Int. Symp. Inf. Theory}, Nice,
France, June 2007.
\bibitem{maxwell-constr}
C.~Measson, A.~Montanari, and R.~Urbanke, ``{Maxwell} construction: The hidden
  bridge between iterative and maximum a posteriori decoding,'' Jun. 2005,
  {arXiv:cs.IT/0506083}.
\end{thebibliography}
%


\end{document}